\documentstyle[11pt,newpasp,twoside]{article}
\input epsf.sty
\index{Pulsars!Radio}
\index{pulsars!B0329+54}
\index{Pulsars!Viewing geometry}
\index{Pulsars!Emission Height}
\reversemarginpar

\begin{document}
\title{Reception of Radio Waves from Pulsars}
 \author{R. T. Gangadhara}
\affil{Indian Institute of Astrophysics, Bangalore -- 560034, India}
\begin{abstract}
The beamed emission by relativistic sources moving along the magnetic dipolar 
field lines occur in the direction of tangents to the field lines. To receive 
such a beamed radiation line-of-sight must align with the tangent within the beaming
angle $1/\gamma,$ where $\gamma$ is the particle Lorentz factor. By solving the 
viewing geometry, in an inclined and rotating dipole magnetic field, we show that
at any given pulse phase observer can receive the radiation only from the specific altitudes.
We find the outer conal emission is received from the higher altitudes than the 
inner conal components including the core. At any pulse phase, low frequency emission
comes from the higher altitudes than the high frequency emission. As an application of our model, we have applied it to explain the emission heights of conal components in PSR~B0329+54.
 \end{abstract}
\section{Introduction}
Pulsar radio emission beam has been widely attempted to interpret in terms of 
emission in purely dipolar magnetic field. Gangadhara \& Gupta (2001) have 
estimated the emission heights of different radio pulse components in PSR~B0329+54 
based on  the aberration--retardation phase shift, and the revised estimates  
are given by Dyks, Rudak, \& Harding (2003). Here we solve the viewing geometry 
and estimate the altitudes from which observer can receive the radio waves.
\section{Emission Beam Geometry}
Consider a magnetic dipole situated at the origin with magnetic axis (${\bf \hat  m}$)
inclined by $\alpha$ with respect to the rotation axis $({\bf \hat  \Omega}),$
and rotated by $\phi'$ around ${\bf \hat  \Omega}.$  Let ${\bf \hat n} = 
(\sin\zeta,\,0,\,\cos\zeta)$ be the line of sight, where $\zeta=\alpha+\beta,$ 
and $\beta$ is the line of sight impact parameter.

In a relativistic flow, the emitted radiation is beamed in the direction of
field line tangent ${\bf \hat b}$, so at any instant the observed radiation 
comes from a spot in the magnetosphere where the tangent vector  points
in the direction ${\bf \hat n}$ of observer. For receiving such 
radiation the semi opening angle of emission beam $\Gamma =\arccos({\bf 
\hat n . \hat m})$ must be approximately equal to the opening angle of field 
lines $\tau=\arccos({\bf \hat b . \hat m}).$ Therefore, the magnetic 
{\it colatitude} $(\theta )$ is given by
\begin{equation}
\cos(2\,\theta)=\frac{1}{3}\left(\cos \Gamma\,\sqrt{8 + \cos^2 \Gamma}-\sin^2 \Gamma\right)\:, 
                   \quad\quad -\pi\leq\Gamma\leq\pi\: .
\end{equation}
Next, the magnetic {\it azimuth} $(\phi)$  of the emission point can be obtained by finding $\bf \hat b $ which is parallel to ${\bf \hat n} :$
\begin{equation}
\sin\phi= -\sin \zeta\,\sin \phi'\csc \Gamma  .
\end{equation}
For $\beta >0,$ on leading side the maximum value for $\phi$ allowed by the 
viewing geometry is $\pi/2,$ which in turn allows to find the maximum pulse 
window $W=2\phi',$ where $\phi'$ is the pulse phase at which $\phi$ approaches 
$\pi/2 .$  Using $\alpha=30^\circ$ and $\beta=2.1^\circ$ for PSR~B0329+54, 
we find $\theta_{\rm max}\sim 8^\circ ,$ $\Gamma_{\rm max}\sim 12^\circ$ and 
$W\sim 46^\circ .$ 

Pulsar radio emission is generally believed to be coherent curvature radiation by secondary pair plasma streaming along the dipolar magnetic field lines. 
The curvature emission peaks at the characteristic frequency
(e.g., Eq.~45, Ruderman \& Sutherland 1975). For a given frequency and a 
Lorentz 
factor $\gamma ,$ we can estimate the radius of curvature $\rho$, which in tern 
allows to 
find the field line constant. So, using $\gamma=340$ and 390 we estimated 
the emission height (see, Fig.~1a) of radiation at 325 MHz and 606 MHz, respectively.
On the other hand by accepting the emission heights derived from the 
aberration-retardation phase shift, we estimated $\gamma$ and $\rho$ expected:
$0.15\leq\rho/r_{\rm LC}\leq 0.33,$ $286\leq\gamma\leq 370$ for 325 MHz 
emission, and $0.12\leq\rho/r_{\rm LC}\leq 0.26,$ $328\leq\gamma\leq 420$ for 
606 MHz.

The polar cap with foot location of emission associated field lines is given in
Fig.~1b, where $z-$axis is chosen to be parallel to ${\bf \hat m}$ and x to lie 
in the ${\bf \hat \Omega}$-${\bf \hat m}$ plane. It is nearly elliptical with radius of 
164~m and 171~m in x and y directions.

\begin{figure}
\vskip -0.5 truecm
\plotone{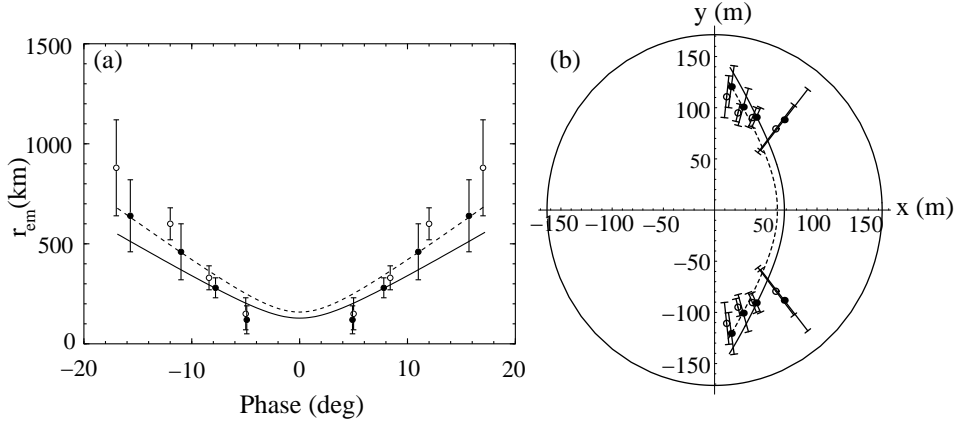}
\vskip -2.0 truecm
\caption{\hskip -0.4 truecm(a) Emission heights in PSR~B0329+54: solid and 
dashed line curves are for the emissions at 606 MHz and 325 MHz, respectively. 
The emission heights estimated from aberration-retardation phase shift are 
superposed: the points marked with $\circ$ for 325 MHz and 
{\large$\bullet$} for 606 MHz. (b) Polar cap with foot of emission associated
field lines.}
\end{figure}

\end{document}